\documentclass[twocolumn,showpacs,preprintnumbers,amsmath,amssymb]{revtex4}
\usepackage{graphicx}% Include figure files
\usepackage{dcolumn}% Align table columns on decimal point
\usepackage{bm}% bold math
\usepackage{amssymb}

\newcommand\nuh{\nu_h}

\newcommand\pair{e^+ e^-}
\newcommand\mix{|U_{\mu h}|^2}
\newcommand\mixt{|U_{\tau h}|^2}

\newcommand\nuhd{\nu_h \rightarrow \gamma \nu}
\newcommand\ee{e^+e^-}
\newcommand\nus{\nu_h}
\newcommand\dstt{D_s^+ \to  \tau^+ \nu_\tau}

\def\address{\@ifstar{\address@star}%
  {\@ifnextchar[{\address@optarg}{\address@noptarg}}}

\begin{document}

\author{S.N.~Gninenko}
\affiliation{Institute for Nuclear Research, Moscow 117312}

%\preprint{APS/123-QED}

%\title{Manuscript Title:\\with Forced Linebreak}% Force line breaks with \\
\title{Sterile  neutrino decay as a common origin for\\
 LSND/MiniBooNe and T2K excess events}
%\title{The  MiniBooNE anomaly and heavy neutrino decay}

\date{\today}% It is always \today, today,
%\date{June 17, 2009}% It is always \today, today,

\begin{abstract}
We point out that the excess of electron-like neutrino 
events recently observed by the T2K collaboration may have a common origin with the 
similar excess events previously reported by the LSND and MiniBooNE experiments and interpreted 
as a signal from the radiative decays of a sterile neutrino $\nuh$ with the mass around 50 MeV
produced in $\nu_\mu$  neutral current (NC) interactions. In this work  
we  assumed that the $\nuh$ can also be produced in $\nu_\tau$NC reactions.

\end{abstract}

\pacs{14.80.-j, 12.60.-i, 13.20.Cz, 13.35.Hb}% PACS, the Physics and Astronomy

\maketitle

%%%%%%%%%%%%%%%%%%%%%%%%%%%%%%%%%%%%%%%%%%%%%%%%%%%%%%%%%%%%%%%%%%%%%%%%%%%%%%%

%\section{Introduction}
 
Over the past 10 years there is a  puzzle of the 3.8 $\sigma$ event excess observed 
by the LSND collaboration \cite{lsndfin}. 
This excess originally interpreted as a signal from 
$\overline{\nu}_\mu \to \overline{\nu}_e$ 
 oscillations was not confirmed by further measurements from the similar KARMEN experiment
\cite{karmen}.  
The MiniBooNE experiment, designed to examine the LSND effect, 
 also  found no evidence  for $\nu_\mu \to \nu_e$ oscillations.
However,  an anomalous  excess of low energy electron-like (e-like) events
in  quasi-elastic neutrino events has been observed \cite{ minibnu2}.
New MiniBooNE  results from a search for
$\overline{\nu}_\mu \to \overline{\nu}_e$ oscillations  also
 show an excess of events, which has a small probability to be identified as  the 
background-only events \cite{minibnub}. The data are found to be  consistent with $\overline{\nu}_\mu \to \overline{\nu}_e$ oscillations in the 0.1 eV$^2$
range and with the evidence for antineutrino oscillations from the LSND.

In the recent work \cite{sngprd1} (see also \cite{sngprl, gg, sngprd2}) it has been shown  that puzzling LSND, KARMEN and MiniBooNE results could all be explained in a consistent way  by assuming  
the existence of a  heavy sterile neutrinos ($\nu_h$). The $\nu_h$ is created in  
$\nu_\mu$ neutral-current interactions and decay subsequently   into  
a photon and a lighter  neutrino $\nu$ in the  LSND and MiniBooNE detectors, 
but it cannot be produced in the KARMEN experiment  due to the high energy threshold.
The  $\nu_h$ could be Dirac or Majorana type, and it could be produced, e.g. through the $\nu_\mu - \nu_h$ mixing.
  The $\nu_h$ could decay  
{\em dominantly} into $\gamma \nu$ pair  if, for example,  there is a large enough  transition
magnetic moment between the $\nu_h$ and $\nu$ mass states.  
Such kind of $\nu_h$'s may be present in many interesting extensions of 
the standard model, see e.g. \cite{moh}. 
Assuming the $\nu_h$ is produced through mixing with $\nu_\mu$, 
the combined analysis of the LSND and MiniBooNe excess events suggests that 
 the $\nu_h$  mass, mixing strength, and lifetime are,  respectively,  in the range
 \begin{eqnarray}
 40\lesssim m_h \lesssim 80~ \text{ MeV},~ 10^{-3}\lesssim |U_{\mu h}|^2 \lesssim 10^{-2}, \nonumber \\
 10^{-11}\lesssim \tau_h\lesssim 10^{-9}~s.
 \label{param}
 \end{eqnarray}
A detailed discussion of consistency of these values  with the constraints from
previous searches for heavy neutrinos \cite{pdg} as well as 
 of the interpretation of the $\nuh$ decay in terms of transition magnetic moment is presented in \cite{sngprd1, sngprd2}.
 Briefly, the mixing of \eqref{param} is not constrained by the limits from  
  the most sensitive experiments searched for extra peaks in two-body $\pi, K$ decays \cite{pdg}, because  
the $\nu_h$ mass range of \eqref{param} is 
 outside of the kinematical limits for $\pi_{\mu 2}$ decays, and  not accessible to 
 $K_{\mu 2}$ experiments due to experimental resolutions.
The parameter space of \eqref{param} cannot be ruled out by the results of high energy  neutrino experiments, 
such as  NuTeV or CHARM, as they searched for $\nu_h$'s of higher   masses ($ m_h \gtrsim 200~ \text{MeV}$)  
 decaying into muonic  final states ($\mu \pi \nu,~\mu \mu \nu,~\mu e \nu, ..$) \cite{pdg}, which are not 
 allowed in our case. The best limits on $|U_{\mu h}|^2$  derived for the mass range \eqref{param} 
  from the search for $\nu_h \to e^+ e^- \nu$ decays in the PS191 experiment \cite{ps191}, 
 as well as the LEP bounds \cite{aleph},  are found to be compatible with \eqref{param} 
 assuming  the dominance of the $\nuh$ decay. New limits on  mixing $|U_{\mu h}|^2$  obtained by 
 using the recent results on precision measurements of the muon Michel parameters by the TWIST 
experiment \cite{twist} are also found to be consistent with \eqref{param}.
 Finally, the most stringent 
bounds on $\mix$ coming from the primordial nucleosynthesis and SN1987A considerations, 
as well as the limits from the atmospheric neutrino experiments,
are also evaded due to the short $\nu_h$ lifetime.

Very recently, the T2K  collaboration, which study  $\nu_\mu$ neutrino 
neutrino interactions in a long baseline experiment at J-PARK, 
 has reported on observation of an excess of  electron-like events
in charge-current quasi-elastic (CCQE) neutrino events  
over the expected standard neutrino interaction events  \cite{t2k}. 
A confirmation of the T2K excess  and clarification of its origin 
have great importance for neutrino physics.  Although the most popular 
mechanism for this excess is $\nu_\mu \to \nu_e$ neutrino 
 oscillations with  nonzero value of the neutrino mixing angle $ \Theta_{13}$, 
 one can still reasonable  ask if neutrino oscillations are the only explanation for the T2K result, see e.g \cite{gibin}.
In this work we study a possible manifestation of the presence 
of $\nu_h$'s in the J-PARC neutrino beam and show that 
the excess of e-like events observed by T2K could be interpreted as a signal 
from the  production and radiative decay of a $\nu_h$  previously suggested for the explanation of    
the origin of similar  excess events observed by the LSND and MiniBooNe experiments.

In the T2K experiment, specifically designed to search for $\nu_\mu \to \nu_e$ oscillations 
and for measurements at the first $\nu_\mu - \nu_\tau$ oscillation maximum (corresponding to the atmospheric neutrino parameters sin$^2 2\Theta_{23}=1$ and $\Delta m_{23}^2 = 2.4\times 10^{-3}$ eV$^2$), the  neutrino flux at the far detector location  is dominated  by $\nu_\tau$'s. 
Therefore,  we will assume in the following that the $\nuh$ can be also produced in $\nu_\tau$NC 
interactions, e.g. through the mixing with the $\nu_\tau$ neutrino \cite{ref}. 
Taking into account that the corresponding mixing strength $\mixt$ is purely 
constrained by existing experimental data, makes this assumption more interesting.
 For example, the bounds from  the NOMAD experiment 
 $\mixt \lesssim (4 - 1)\times 10^{-2}$ for the $\nuh$ masses from  40 to  80 MeV  were  obtained 
 from the search for the $\nu_h \to \nu_\tau \ee$ decay under assumption  that this decay is the principal decay mode of the $\nu_h$. These limits can be significantly relaxed assuming the dominance of the $\nuhd$ decay. Direct searches for radiative decays of an 
excited neutrino $\nu^* \to \nu \gamma$ produced in $Z\to \nu^* \nu$ decays have also been performed at LEP
\cite{pdg}. For the best  limit Br$(Z\to \nu^* \nu)$Br$(\nu^* \to \nu \gamma)< 2.7\times 10^{-5}$
from ALEPH \cite{aleph}, taking into account  
$\frac{Br(Z\to \nu \nu_h)}{Br(Z \to \nu \nu )} \simeq  |U_{\tau h}|^2$, we find \cite{sngprd1}
\begin{equation}
|U_{\tau h}|^2\times \frac{m_{\nu_h}[\text{MeV}]}{\tau_{\nu_h}[s]} < 4.8 \times 10^{9}. 
\label{aleph}
\end{equation}
For the mass and mixing range of \eqref{param} it results  in $\tau_{\nu_h} \gtrsim 10^{-11}$ s.  

In addition, 
there is also  a hint from the measurements of the $\dstt$ decay rate  at CLEO~\cite{alex,ony}.
% It should be mentioned that the best precision results in measurements of the decay
%rate  of  $\dstt$  achieved at CLEO~\cite{ony} are 
%consistent with theoretical predictions. However,  from other
%based on study of some decay mode of outgoing$\tau$-lepton,
It has been observed that  for some decay modes of outgoing $\tau$-leptons,
the obtained decay rate  of $\dstt$ is significantly higher than the predicted one,  so
that combined branching ratio deviates from the theoretical prediction 
at the level of $\simeq 10\%$.
One may speculate that this  
inconsistency  is due to existence of a sterile neutrino mixed into the tau neutrino. This would 
result in an additional contribution to the $\dstt$ decay rate from the decay $D_s^+ \to  \tau^+ \nu_h$.
It has been found that for the mass range below 190 MeV 
the mixing strength required to explain $\dstt$ discrepancy 
should be $|U_{\tau h}|^2 = 0.16 \pm 0.09$  \cite{gg}, which for the mass and lifetime range of 
\eqref{param}  is consistent with \eqref{aleph} within the large error.
 
The T2K experiment is described in details in \cite{t2kdet}. 
It uses an almost pure off-axis  $\nu_\mu$ beam originated from
the  $\pi^{+}$ and $K$ decays in flight, which are  generated 
  by 30 GeV protons from the J-PARC Main Ring accelerator.
The detector consists of a near detector complex, used to measure 
precisely the $\nu_\mu$ flux and to predict the standard neutrino interaction rate in the far detector, which is 
the Super-Kamiokande (SK) water Cherenkov detector located at the distance of 295 km from the proton target.
The SK detector is a cylindrical tank,  about  40 m  in  diameter and  40 m height, 
 filled with $\simeq$50 kt of purified water \cite{sk}. The detector has a fiducial volume (FV) of 22.5 kton
 within its cylindrical inner detector (ID), and a 2 m-wide  outer detector (OD) served as a veto against
 cosmic rays and neutrino interactions in the surrounding rock. 
The Cherenkov light rings generated by  muon, electron and converted photon 
tracks are used for the reconstruction of the events. 
The T2K search for e-like events from $\nu_\mu \to \nu_e$ neutrino oscillations  
uses the data sample collected during the years 2010-2011 \cite{t2k}.
The strategy of the analysis is to identify the $\nu_e$CCQE candidate events  
by reconstructing in the FV isolated single e-like rings 
 that are accompanied  by no other activity in the outer detector.\ The measured rate 
of the e-like events is then compared to the one  expected from known reactions.
\begin{figure}[tbh!]
\includegraphics[width=0.35\textwidth]{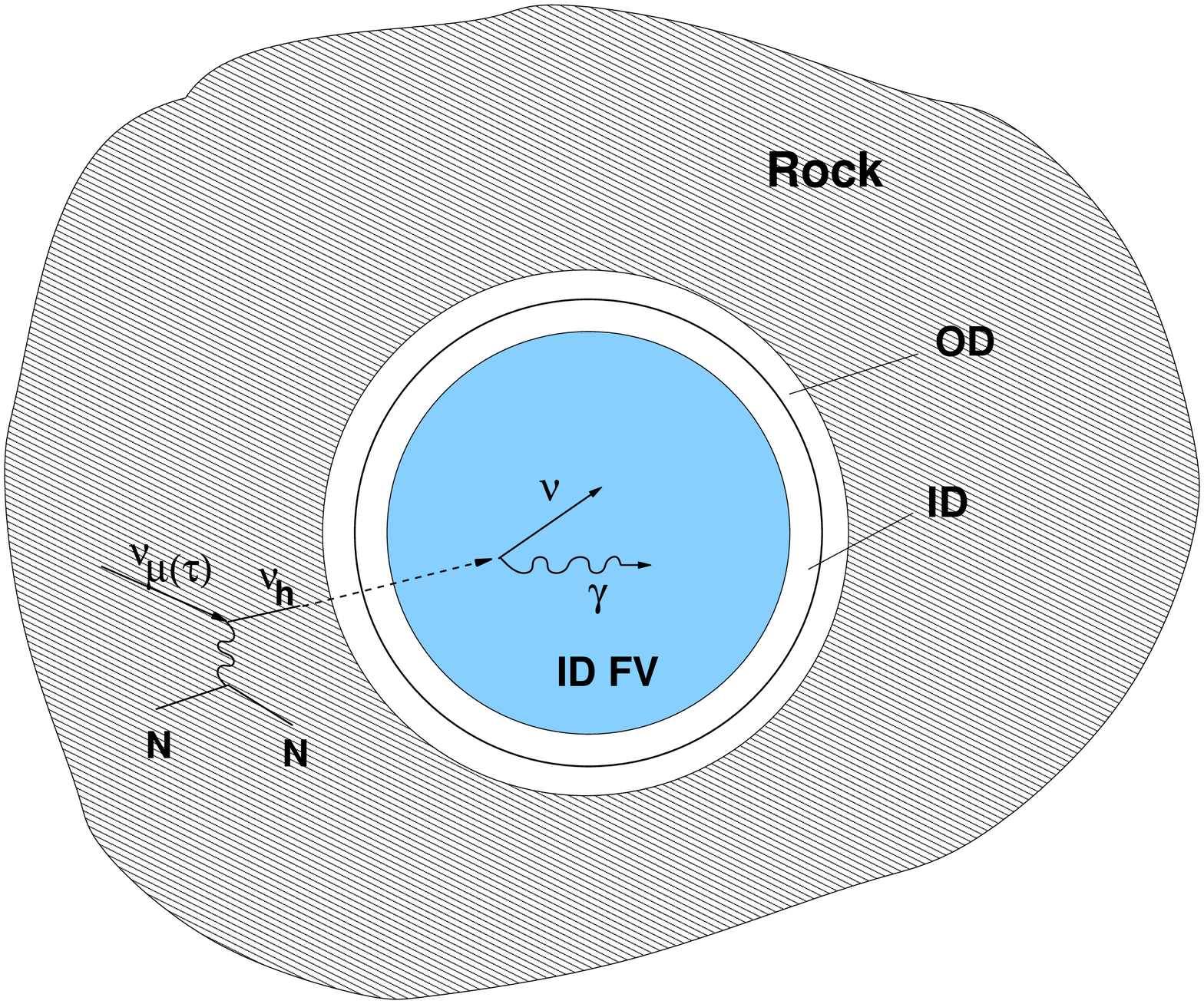}% Here is how to import EPS art
\caption{\label{fig:diag} Schematic illustration of the production and 
subsequent radiative decay of a heavy neutrino in the SK detector (top view). 
    The $\nu_h$'s are  produced  in $\nu_{\mu(\tau)}$NC interactions 
    of the J-PARC $\nu_\mu$ neutrino beam either in the FV, the ID region outside FV, 
    the OD or, as shown, in the surrounding rock. In the later case the $\nu_h$  would penetrate the rock shield and would be observed 
 in the SK neutrino detector through their $\nuhd$  decays followed by the decay  
photon conversion into an $e^+e^-$ pair in the ID FV. }
\label{sk}
\end{figure}  
 An excess of $\Delta N=4.5$ electron-like events (6 events observed and 1.5$\pm$0.3 expected)  has been observed in the data   accumulated with $1.43\times 10^{20}$ protons on target (pot).
For the following discussion several distinctive features of the excess events
are of importance \cite{t2k}: a) the excess is  observed for single e-like tracks, 
originating either from an 
electron, or from  a photon converted into a $\pair$ pair with a typical 
opening angle $\simeq m_e/E_{\pair}< 1$ degree (for
 $E_{\pair} > 100$ MeV), which is 
too small to be resolved into two separate Cherenkov rings in the SK detector(here,
$m_e, E_{\pair}$ are the electron mass and the $\pair$ pair energy);
b) the reconstructed neutrino 
energy is in the range $200 < E^{QE}_\nu < 1000$ MeV. 
The variable  $E^{QE}_\nu$  is calculated under 
the assumption that the observed  electron track originates from a 
$\nu_e$CCQE  interaction;  c) the visible 
energy $E_{vis}$   is required to be  
$E_{vis} \gtrsim 100$ MeV;
d) the angular distribution of 
the excess events with respect to the incident neutrino direction 
is wide and consistent with the shape expected 
from $\nu_e$CCQE interactions; e) there is  no additional significant activity in the OD detector.
 
To satisfy the criteria a)-e), we propose that  the excess events are  
originated from the production and subsequent radiative decay of a heavy neutrino $\nu_h$
in the FV of the SK detector.
The heavy neutrinos are assumed to be produced in the neutral-current quasi-elasic (NCQE) interactions
$\nu_{\mu(\tau)} + N \to \nu_h +N$
 of  muon or tau neutrinos either in the  SK FV,  the ID region outside FV, OD region, or in the surrounding rock.  
 In the later case, if $\nu_h$ is a relatively long-lived particle, the flux of $\nu_h$'s
would penetrate  the  rock shielding without significant attenuation  and would be observed in SK
through their $\nuhd$ decays with the subsequent conversion $ \gamma \to \pair$
 of the decay photons in the SK water target, as  schematically illustrated  in Fig. \ref{sk}.
 Similar to $\nu_\mu \to \nu_e $ neutrino oscillations, the occurrence of $\nuhd$ decays would appear as 
an excess of single e-like events  from decay photon conversion in the SK  detector,    
above those expected from standard neutrino interactions.
To make a quantitative estimate, 
we performed simplified simulations of the $\nu_h$ production 
and decay processes in the SK discussed above.

 The flux of the  produced $\nu_h$'s can be calculated by using the following equation for the $\nu_h$ production  cross section, 
\begin{equation}
 \sigma(\nu_{\mu(\tau)} N \to \nu_h  N) = \sigma(\nu_{\mu(\tau)} N \to \nu_{\mu(\tau)} N) |U_{\mu(\tau) h}|^2 f,  
\label{crossec}
\end{equation}
 where $\sigma(\nu_{\mu(\tau)} + N\rightarrow \nu_{\mu(\tau)} + N)$ is the 
cross section for $\nu_{\mu(\tau)}$NCQE interactions and  $f$ is the  phase space 
 factor calculated for this two-body reaction, which takes into account  dependence on the $\nu_h$ mass. The  energy spectra of the produced $\nu_h$'s, whose momenta  pointing to 
the SK fiducial volume,  as well as the angular distribution of the  $\nu_h$'s,
 were calculated for different $\nu_h$ masses 
 by taking into account the $\nu_\mu, \nu_\tau$ energy distributions at the far detector \cite{t2k}.
In these simulations  we used a parametrized $\nu_\mu$ energy spectrum obtained at far detector from the  
reconstructed $\nu_\mu$CCQE events \cite{t2k}.
 The $\nu_\tau$ energy distribution at the FD position was calculated from the primary $\nu_\mu$ spectrum
within the standard two-neutrino oscillations scheme for the atmospheric parameters 
quoted above.  
%The $\nu_\tau$ beam is peaked  around $\sim 600$ MeV, has a mean energy of  
% $\sim 800$ MeV and a high energy tail up to $\sim$ 5 GeV.
 The total number of NC events in the FV of the SK  detector was used for 
normalization. 
 \begin{figure}[tbh!]
\includegraphics[width=0.5\textwidth]{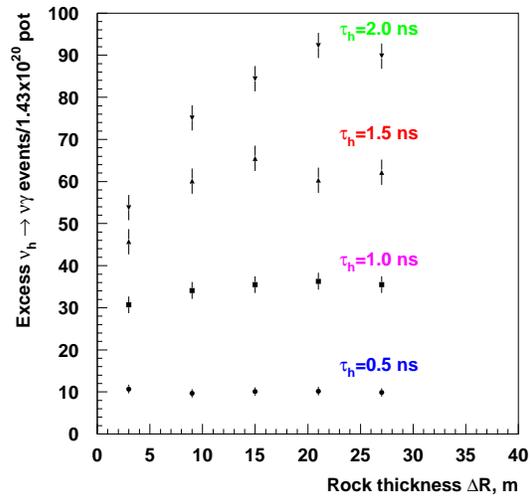}% Here is how to import EPS art
\caption{ The number of $\nuh$ decays from the $\nu_h$'s
 produced in the rock  as a function of rock thickness calculated for $\nu_\mu$ and $\nu_\tau$
 energy spectra at the far detector, assuming $\mix= 1$ and $\mixt= 1$, for different $\tau_h$ values 
 shown in the plot, $a=0$, and $1.43\times 10^{20}$ pot. The fraction of $\nuhd$ decays from $\nu_\mu$NC 
 reactions is $\lesssim 15 \%$, while the rest is due to $\nu_\tau$NC interactions.}
\label{rock}
\end{figure}  

Once the $\nus$ flux was known, the next step was to calculate the $\pair$ spectrum based on  
the  $\nuhd$ decay rate. For a given flux  $\Phi_{\nu_h}$, the expected number of signal events from 
$\nuhd$ decays occurring within the fiducial  length $L$ of  the SK 
detector with the FV entrance point  located  at a distance $L'$ from the $\nu_h$ production vertex is given by 
\begin{eqnarray}
n_{\nuh} = \sum_{\nu_\mu,\nu_\tau}\int  A \Bigl(\frac{d\Phi_{\nu_h}}{dE_{\nuh}}\Bigr)_{\nu_\mu(\nu_\tau)}\text{exp}\bigl(-\frac{L'm_{\nu_h}}{p_{\nu_h}\tau_h}\bigr) \\ \nonumber 
\bigl[1-\text{exp}\bigl(-\frac{L m_{\nus}}{p_{\nuh}\tau_h}\bigr)\bigr] 
 \frac{\Gamma_{\gamma\nu}}{\Gamma_{tot}} \varepsilon_\gamma \varepsilon_{\ee}  dE_{\nuh}dV
\label{rate}
\end{eqnarray}
where $p_{\nuh}$ is the $\nu_h$ momentum and  $\tau_h$ is its 
lifetime at the rest frame, $\Gamma_{\pair},~\Gamma_{tot}$
are the  partial and total  mass dependent 
$\nuh$-decay widths, respectively, $\varepsilon$ is the $\pair$ pair reconstruction efficiency and 
the integral is taken over the FV,  ID region outside FV, OD and the surrounding rock volume. 
It is assumed that the total rate $\Gamma_{tot}$ of the $\nu_h$ decays is dominated by the 
 radiative decay $\nuhd$, i.e. the branching fraction    
$BR(\nuhd) = \frac{\Gamma(\nuhd)}{\Gamma_{tot}}\simeq 1$ \cite{sngprd1}.
\begin{figure}[h]
\begin{center}
    \resizebox{9cm}{!}{\includegraphics{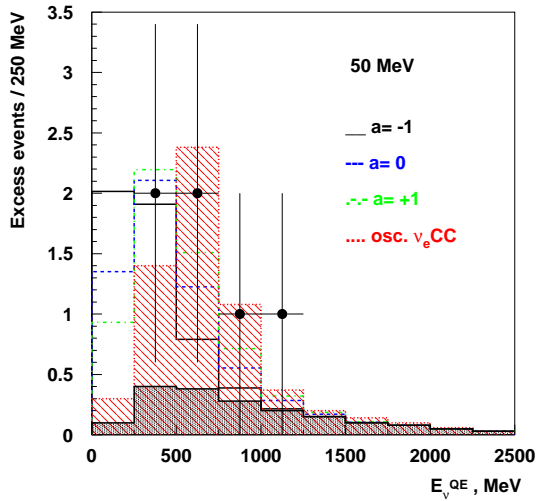}}
     \caption{ Distributions of the excess events reconstructed as $\nu_e \rm{CCQE}$ events  in the SK detector
as a function of variable $E^{QE}_\nu$  for  $E_{vis}> 100$ MeV from the experimental data sample (dots), and from a combination of the  $\nuhd$ decay of $\nuh$'s produced in $\nu_\tau$NC plus expected neutrino background (bottom shaded histogram, from Ref. \cite{t2k})
 calculated for $a= -1$ (solid line), $a= 0$ (dashed line),  and $a= +1$ (dashed-dotted line)  cases shown in the plot,
 the $\nu_h$ mass of 50 MeV,  and the $\nu_h$ lifetime $\tau_{\nu_h}= 10^{-9}$ s.     
  Error bars include only statistical errors. The distributions are normalized to six events, which 
  corresponds $\mixt \simeq 0.025$.
A distribution of neutrino background plus $\nu_\mu \to \nu_e$ neutrino 
oscillations at $\rm{sin}^22\Theta_{13}=0.1$ (dotted histogram, from Ref.\cite{t2k}) is also shown
for comparison.}
\label{eque50}
\end{center}
\end{figure}
The acceptance $A$ of the SK detector was calculated by tracing the 
produced  $\nus$'s to the detector FV  taking  momentum and angular distributions into account.
 The energy of the photon from the $\nuh$ decay depends on the initial 
neutrino energy and on the center-of-mass angle $\Theta$ between the photon 
momentum and the $\nu_h$ momentum  direction. Therefore, the photon laboratory energy spectrum 
depends on the c.m. angular distribution, which is generally given by 
$dN/d\rm{cos}\Theta$ $\simeq 1+ a \cdot \rm{cos}\Theta$, where
asymmetry coefficient $a$ is in the range $-1 < a < 1$ for Dirac, and $a=0$ for Majorana neutrinos \cite{vogel}.  
The reconstruction efficiency of the photon converted in the  fiducial
 volume of the SK detector was taken to be $\simeq 70\%$  from the T2K  simulations 
 of the $\nu_e$ QECC events \cite{t2k}. 
An example of the calculated number of the $\nuh$ decays in the SK FV is shown in Fig. \ref{rock}
as a function of the thickness of the rock surrounging the detector for the  $\nu_h$ mass of 50 MeV 
and several $\tau_h$ values. It is seen, that if  the $\nu_h$ is a relatively short-lived particle, 
i.e $\frac{L m_{\nu_h}}{p_{\nu_h} \tau_{h}} < 1 $ the number of the signal events is quickly saturated
and does not depend on the rock thickness.
 Neutrino interactions, with little hadronic activity in the final state, occuring in the OD  or SK support structure,  as well as in the  part of the ID outside FV, can also  yield an isolated e-track from the $\nuh$ decay in the FV.
\begin{figure}[h]
\begin{center}
    \resizebox{10cm}{!}{\includegraphics{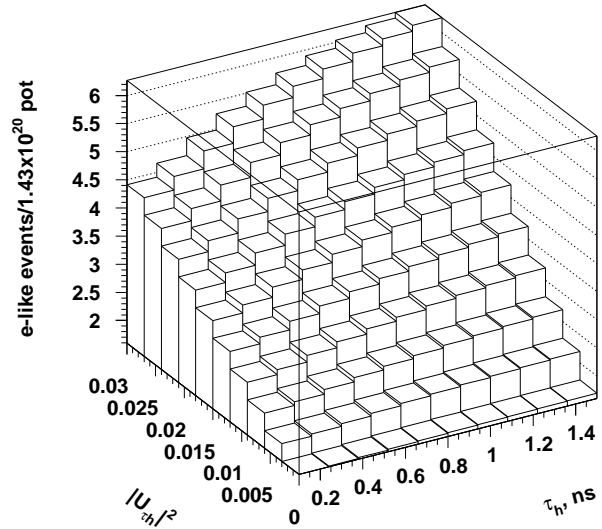}}
     \caption{ Number of expected e-like events from $\nuh$ decays plus neutrino background in the SK FV for $1.43\times 10^{20}$ pot as a function of the mixing strength $|U_{\tau h}|^2$ and the $\nu_h$ lifetime $\tau_h$ from the region of \eqref{param} calculated for $m_h = 50$ MeV and $a=0$.}
\label{biplot}
\end{center}
\end{figure}
The attenuation of the $\nu_h$ - flux due to $\nu_h$  interactions 
in the rock with the average density 3.2 $g/cm^3$ was found to be negligible.
 
 In Fig. \ref{eque50} an example of distributions of 
the kinematic variable $E^{QE}_\nu$ for the excess $\nuh$ decays events in the SK detector
 reconstructed as $\nu_e$CCQE events  plus neutrino background predicted in \cite{t2k} are shown
 for $E_{vis}> 100$ MeV, $m_{\nu_h} = 50$ MeV,  $\tau_{\nu_h}= 10^{-9}$ s, and 
 different values of $a$. 
  These distributions are calculated for the  dominant production of $\nuh$'s 
  by $\nu_\tau$'s   assuming 
 that the  $\pair$ pair from the converted photon 
 is mis-reconstructed  as a  single track from the $\nu_e$CCQE reaction.
 The distributions are then normalized to six events, which corresponds 
 $\mixt \simeq 0.025$, to compare them with the T2K data.
Simulations are in reasonable agreement with the experimental distributions.  
For instance, for the distributions shown in Fig. \ref{eque50} the $\chi^2$ test of their  consistency 
with T2K data yields  $p-values$ of 0.79, 0.87, and 0.92 for  $a=-1,~ 0$, and $+1$, respectively. In  
these calculations only statistical errors  for both experimental, as reported in  \cite{t2k}, and 
 simulated spectra  are included. The test is based on the 
 method of comparison of experimental and simulated histograms recently proposed in \cite{comp},
   which can be apply for analyzing data samples containing small numbers of events per bin.
The simulated excess events, shown in Fig. \ref{eque50}, 
are  mainly distributed  in the region $ 200 \lesssim E^{QE}_\nu \lesssim 1200$ MeV.
The simulations showed that the shape of the $E^{QE}_\nu$  
distributions is sensitive to the choice of the $\nu_h$ mass, $\tau_{\nu_h}$, and, in particular, the $a$-parameter 
values: the smaller the $a$, the softer the spectrum. The distribution of cosine 
of the opening angle between the e-like ring and neutrino beam direction is 
found to be consistent with $\nu_e$ CCQE events.    
Taking into account  the estimated  number of  71 events  expected to be observed in the FV
from the ordinary NC interactions\cite{t2k}, the total number of $\nuh$ events inside 
the fiducial volume of the SK detector is given by  
$n_{\nuh} \simeq n_\mu\mix + n_\tau  \mixt$, 
where coefficients $n_\mu, n_\tau$ vary in the range 
$n_\mu \simeq 12-30$ and $n_\tau \simeq 70-160$,  depending on $\tau_h$. 
As $\mix < \mixt$ by a factor of a few,  mainly  $\nu_\tau$NC interactions in the FV and in the rock contribute to  the total number of excess events. 
The number of excess events from interactions in the  FV is, roughly, 
$\propto \mixt$, while the number of events from the  rock is $\propto \mixt \tau_h$. 
In Fig. \ref{biplot} an example of distribution for the expected number of signal plus neutrino background  events
 is shown in more details in the ($\tau_h; |U_{\tau h}|^2$) parameter space.
For the larger mixing the number of $\nuh$ events increases, while  for  smaller lifetime values it
decreases  due to the more rapid decays of $\nu_h$'s,  and mostly NC interactions in the FV 
contribute in this case.
\begin{figure}[h]
\begin{center}
    \resizebox{10cm}{!}{\includegraphics{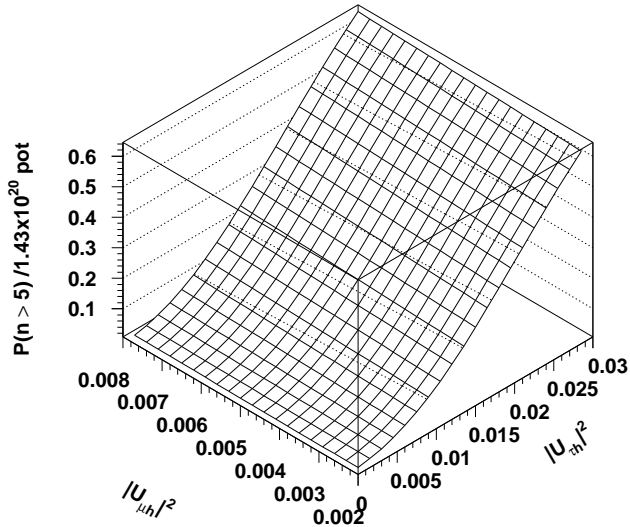}}
     \caption{ The probability  to observe more than 5  events in the T2K experiment
     as a function of $\mix$ and $\mixt$ for $\tau_h = 10^{-9}$ s.}
\label{bimix}
\end{center}
\end{figure}
 The fraction of excess events expected to be seen in the OD depends on the $\nu_h$ lifetime
 and typically is  $\lesssim 20\%$ ($\lesssim$ 1 event) which is consistent with observations \cite{t2k}. It is also interesting to compare 
the spatial distributions of the excess events,  which,  in the case of $\nu_h$ decays, is a combination 
of the uniform distribution from the NC interactions in the FV and a distribution 
from NC interactions outside the FV. 
 The later is the $\nu_h$ lifetime dependent: for 
shorter $\tau_h$ the excess events are expected to be distributed presumably near the edge of the FV.  
One interesting feature of the spatial distribution 
of events in SK is that they are all in the first half of the detector.  
To reproduce this feature one could try to use the $\nu_h$ lifetime shorter than $\tau_h \lesssim 0.5$ ns,
 so that $\nuh$'s produced in  the rock  decay mainly in the first part of the SK. 
 For shorter lifetimes, as one can see from Fig. \ref{rock}, the number of events from the rock 
decreases, and mixing  values  $\mixt \gtrsim 0.03$ are required  
to keep the signal in the range of 4-5 events. Note, that the increase of $\mixt$  results also 
in increase of the relative fraction of  events produced in NC interactions in the FV (typically, about 40\% for the long $\tau_h$'s),  which are distributed uniformly over the FV.
  
The probability $P(n_{\nuh} >5)$ to observe more than 5  events in the T2K experiment calculated as a function 
of mixings $\mix$ and $\mixt$ is  shown in Fig. \ref{bimix}. It was obtained by using the  approach of Ref.\cite{bk} and  taking into account the uncertainties in the background 
estimate \cite{t2k}.
For example, for  $P(n_{\nuh} >5) > 0.25$, using \eqref{aleph} to constrain the $\tau_h$,  the most suitable values of the parameters are 
\begin{eqnarray}
 40\lesssim m_h \lesssim 80~ \text{MeV},~ 10^{-3}\lesssim |U_{\mu h}|^2 \lesssim 10^{-2}, \nonumber \\
10^{-2}\lesssim |U_{\tau h}|^2 \lesssim 3\times 10^{-2},~ 10^{-10} \lesssim \tau_h \lesssim  10^{-9}~{\rm s}. 
 \label{param1}
\end{eqnarray}
In summary, in this work we study  a possible manifestation of the presence 
of heavy neutrinos in the J-PARC neutrino beam and show that, assuming the $\nuh$ mixing into the $\nu_\tau$, 
the T2K excess events  
could originate from the same mechanism as those observed by the LSND and MiniBooNE experiments, namely
 from the production and radiative decay of a sterile  neutrino with properties of \eqref{param1}. 
This interpretation is  found to be 
compatible with all the constraints a)-e). The distribution of the excess events  
in kinematic variable $E^{QE}_\nu$ is found to be consistent with
 of the shapes of  distributions obtained within this interpretation. 
 Our analysis  may be improved by more accurate
and detailed simulations of the T2K experiment, which are
beyond the scope of this work. 
A definite conclusion on the presence of $\nuh$ events can be drawn 
when the T2K statistics is substantially increased. 
Finally, note 
that several ideas on  searching for $\nuh$ in  $\mu$ decays \cite{sngprd2},  with existing neutrino data
\cite{sanjib}, in  radiative $K$ decays \cite{koval, duk}, 
and with neutrino telescopes \cite{masip} have been recently proposed.
 The author thanks D.S. Gorbunov, N.V. Krasnikov and M.E. Shaposhnikov  
  for useful discussions and/or comments, and S.I. Bityukov, A.E. Korneev and D. Sillou for help in calculations.


\begin{thebibliography}{99}
\bibitem{ref}
The suggestion to consider possible mixing with $\nu_\tau$ was proposed to the author by an 
anonymous referee.
\end{thebibliography}
\end{document}